\documentclass[%
print,%单栏 同时下面应改为{revtex4}
superscriptaddress,  %此处为作者单位
amsmath,amssymb,11pt,%字号
]{revtex4}
\usepackage{amsthm}
\usepackage{charter}%此指令是令文章选用某字体.
\usepackage{graphicx}% Include figure files
\usepackage{dcolumn}% Align table columns on decimal point
\usepackage{bm}% bold math
\usepackage{dsfont}
\usepackage[landscape,papersize={297.1mm,210mm},left=1.9cm,right=1.6cm,top=2.7cm,bottom=2.8cm]{geometry}% 此指令是规定A4纸的页边距
\usepackage[                   %dvipdfm, pdflatex,pdftex这里决定运行文件的方式不同
            pdfstartview=FitH,
            colorlinks, %注释掉此项则交叉引用为彩色边框(将colorlinks和pdfborder同时注释掉)
            pdfborder=001,   %注释掉此项则交叉引用为彩色边框
            linkcolor=blue,
            anchorcolor=blue,
            citecolor=blue,
            urlcolor=blue
            ]{hyperref}
\begin{document}
\title{The verification of a requirement of entanglement measures }

\author{Xianfei Qi}
\affiliation {School of Mathematics and Statistics, Shangqiu
Normal University, Shangqiu 476000, China}

\author{Ting Gao}
\email{gaoting@hebtu.edu.cn}
\affiliation {School of Mathematical Sciences, Hebei
Normal University, Shijiazhuang 050024, China}

\author{Fengli Yan}
\email{flyan@hebtu.edu.cn}
\affiliation {College of Physics,
Hebei Key Laboratory of Photophysics Research and Application, Hebei
Normal University, Shijiazhuang 050024, China}

\begin{abstract}
The quantification of quantum entanglement is a central issue in quantum information theory. Recently, Gao \emph{et al}. ( \href{http://dx.doi.org/10.1103/PhysRevLett.112.180501}{Phys. Rev. Lett. \textbf{112}, 180501 (2014)}) pointed out that the maximum of entanglement measure of the permutational invariant part of $\rho$ ought to be a lower bound on entanglement measure of the original state $\rho$, and proposed that this argument can be used as an additional requirement for (multipartite) entanglement measures. Whether any individual proposed entanglement measure satisfies the requirement still has to prove. In this work, we show that most known entanglement measures of bipartite quantum systems satisfy the new criterion, include all convex-roof entanglement measures, the relative entropy of entanglement, the negativity, the logarithmic negativity and the logarithmic convex-roof extended negativity. Our results give a refinement in quantifying entanglement and provide new insights into a better understanding of entanglement properties of quantum systems.
\end{abstract}


\maketitle

\section{Introduction}
As one of the most distinctive features of quantum mechanics, quantum entanglement plays a central role in quantum information processing \cite{RMP81.865}. It has been successfully identified as a key resource in various areas such as quantum computation \cite{Nature2000}, quantum cryptography \cite{PRL67.661}, quantum teleportation \cite{PRL70.1895,EPL84.50001}, quantum dense coding \cite{PRL69.2881} and quantum metrology \cite{Nature2010}.

One of the fundamental problems in quantum entanglement theory is to determine which states are entangled and which are not. For bipartite systems, there are many famous separability criteria \cite{RMP81.865,PR474.1}.  The situation becomes complicated when the number of parties increases. For example, we have to face the problem of $k$-partite entanglement or $k$-nonseparability for a given partition and an unfixed partition in multipartite quantum systems. However, there have been significant advances to construct efficient separability criteria for multipartite systems \cite{QIC2008,QIC2010,PRA82.062113,EPL104.20007,PRA91.042313,SR5.13138,PRA93.042310,SciChina2017}.

On the other hand, quantification of quantum entanglement is also a significant problem in quantum entanglement theory. There are many inequivalent approaches to quantify entanglement, one of which is the axiomatic approach. One can list a set of natural properties that a function ought to satisfy in order to quantify the resource character of entanglement \cite{RMP81.865,QIC2001,QIC2007}. The most basic requirement is that any measure of entanglement $E$ must not increase on average under local operations and classical communication (LOCC) operations. It means that $E$ is a so-called \textit{entanglement monotone} \cite{JMO2000}. There are other desirable properties that are imposed as additional requirements for entanglement measures such as convexity, additivity, \textit{et al}. For bipartite systems, considerable effort has been spent on developing many different entanglement measures \cite{QIC2001,QIC2007,JPA47.424005}. However, the promising results in quantifying bipartite entanglement did not easily generalize to systems of more parties owing to the complicated structure of multipartite entangled states. Despite the difficulty, a number of computable entanglement measures \cite{PRA68.042307,PRL93.230501,PRA83.062325,PRA86.062323} quantifying multipartite entanglement have been presented.

In this paper we focus on the refinement of axiomatic approach of quantifying entanglement. The concept of \textquotedblleft the permutationally invariant (PI) part~$\rho^{\text{PI}}$ of a density matrix $\rho$\textquotedblright~is very useful for both efficient quantum state estimation and entanglement characterization of multipartite systems. In \cite{PRL112.180501}, the authors argued to add as requirement on any (multipartite) entanglement measure $E$ that it satisfies $E(\rho)\geqslant \max\limits_{\text{all base}~B}E(\rho^{(\text{PI})_{B}})$, where the maximum value is taken over all possible  single-particle basis changes, and gave the pioneering proof by using a quantitative measure of multipartite entanglement called $k$-ME concurrence \cite{PRA86.062323}. However, it is still an open question whether the new requirement is valid for all entanglement measures due to the fact that numerous entanglement measures are constructed based on different approaches. Inspired by the abovementioned work, we attempt to answer the question for bipartite quantum systems. We first present a explicit proof that the new requirement is satisfied by all measures constructed by means of a convex roof, such as entanglement
of formation \cite{PRA54.3824}, geometric measure of entanglement \cite{JPA34.6787}, convex-roof extended negativity \cite{PRA68.062304}, generalized entanglement concurrence \cite{PRA71.012318}, etc. Then we also verify that this new requirement is also satisfied by the entanglement measures which are not based on the convex-roof extension by providing several specific examples.

The organization of this article is as follows: In Sec.~\uppercase\expandafter{\romannumeral 2} we review basic notions and definitions which will be used in the rest of the paper. In Sec.~\uppercase\expandafter{\romannumeral 3}, we prove that most existing entanglement measures of bipartite quantum systems satisfy the new axiom proposed in \cite{PRL112.180501}. Finally, a brief summary is given in Sec.~\uppercase\expandafter{\romannumeral 4}.

\section{Preliminaries}
Before we present our main results, it is meaningful to introduce some necessary knowledge. Throughout the paper, we consider a bipartite $d\times d$-dimensional Hilbert space $\mathcal{H}=\mathcal{H}_{A}\otimes \mathcal{H}_{B}$ shared by $A$ and $B$. For simplicity, we assume that the dimensions of two subsystems are equal, but this is not essential.

 The permutationally invariant part of a bipartite quantum state $\rho$ is defined as
\begin{equation}
\rho^{\text{PI}}=\dfrac{1}{2}\sum\limits_{n=1}^{2}\Pi_{n}\rho\Pi_{n}^{\dag},
\end{equation}
where $\Pi_{n}\in \mathcal{S}_{2}$. Here, $\mathcal{S}_{2}$ is the symmetric group of all permutations on $2$ particles. That is, $\mathcal{S}_{2}=\{\Pi_{1}, \Pi_{2}\}$  which $\Pi_{1}$ is the identity permutation, $\Pi_{2}$ acts on $\mathcal{H}$ by means of permuting two particles.

 A real function $f:\mathcal{T}(\mathbb{C}^{d})\longrightarrow R$ on the space $\mathcal{T}(\mathbb{C}^{d})$ of density matrices on $\mathbb{C}^{d}$ is called unitarily invariant concave function \cite{JMO2000} if it satisfies
\begin{equation}
f(U\rho U^{\dagger})=f(\rho)
\end{equation}
for any $\rho\in \mathcal{T}(\mathbb{C}^{d})$ and unitary matrix $U$ on $\mathbb{C}^{d}$, and concavity, i.e.,
\begin{equation}
f[p\rho_{1}+(1-p)\rho_{2}]\geqslant pf(\rho_{1})+(1-p)f(\rho_{2})
\end{equation}
for any $\rho_{1}, \rho_{2}\in \mathcal{T}(\mathbb{C}^{d})$ and $0\leqslant p\leqslant 1$.

Denote by $\mathcal{F}_{uc}$ the set of real unitarily invariant concave functions on the space of density matrices. In \cite{JMO2000}, a one to one mapping between entanglement monotones and real unitarily invariant concave functions was established. Any function $f\in \mathcal{F}_{uc}$ can be used to construct an entanglement monotones $E_f$ on $\mathcal{T}(\mathcal{H})$ in the following way: one starts by defining the function $E_f$ on pure state $|\psi\rangle\in \mathcal{H}$,
\begin{equation}
E_{f}(|\psi\rangle):= f(\text{tr}_{B}[|\psi\rangle\langle\psi|])\qquad (=f(\text{tr}_{A}[|\psi\rangle\langle\psi|]), \label{theorem11}
\end{equation}
and then extends it to mixed states $\rho\in\mathcal{T}(\mathcal{H})$ by  the convex roof method,
\begin{equation}
E_{f}(\rho):= \min\sum\limits_{i}p_{i}E_{f}(|\psi_i\rangle), \label{theorem12}
\end{equation}
where the minimization is taken over all possible pure state ensemble decompositions $\{p_{i}, |\psi_i\rangle\}$ for which $\rho=\sum_{i}p_{i}|\psi_{i}\rangle\langle\psi_{i}|$.

The connection between entanglement monotones and real unitarily invariant concave functions is summarized as \cite{JMO2000,PRA96.032316}: For any $f\in \mathcal{F}_{uc}$, the function $E_f$ defined by (\ref{theorem11}) and (\ref{theorem12}) is an entanglement monotone. Conversely, the restriction to pure states of any entanglement monotone is identical to $E_f$ for certain $f\in \mathcal{F}_{uc}$.

This result provides a very elegant way of constructing many useful convex-roof entanglement measures, for instance, entanglement
of formation \cite{PRA54.3824}, geometric measure of entanglement \cite{JPA34.6787}, convex-roof extended negativity \cite{PRA68.062304}, generalized entanglement concurrence \cite{PRA71.012318}, \textit{etc}. Similar approach can also be used to construct various coherence measures in the resource theory of coherence \cite{QIC15.1307,PRA92.022124,PRL116.120404,QIP16.198,JPA50.285301} (see review papers \cite{RMP89.041003,PR2018}).

\section{Main results}
In this section, we are ready to show the main results of the article. The following Observation, as a part of a result
first established by Gao \textit{et al} \cite{PRL112.180501}, provides a reasonable requirement as well as strong bounds for entanglement measures. We restate the result here and present the detailed proof to verify the validity for the most known entanglement measures of bipartite quantum systems.

\emph{Observation} \cite{PRL112.180501}. For the most known entanglement measures $E$, the maximum of $E$ of the permutational invariant part of $\rho$  is not greater than $E$ of the original state $\rho$. That is,
\begin{equation}
E(\rho)\geqslant \max\limits_{\text{all base}~B}E(\rho^{(\text{PI})_{B}}),\label{observation}
\end{equation}
where $\rho$ is an arbitrary bipartite quantum state and the maximization in the right-hand side is taken over all possible single-particle basis changes.

\emph{Proof}. We first consider the case of convex-roof entanglement measures. Note that $f(\text{tr}_{B}[|\psi\rangle\langle\psi|])=f(\text{tr}_{A}[|\psi\rangle\langle\psi|])$, then for any pure bipartite state $|\psi\rangle$ and permutation $\Pi\in \mathcal{S}_{2}$, it is clear that
\begin{equation}
E_{f}(\Pi|\psi\rangle)=E_{f}(|\psi\rangle).
\end{equation}
According to the definition of entanglement measure \cite{JMO2000}, we have
\begin{equation}
E_{f}((|\psi\rangle\langle\psi|)^{\text{PI}})\leqslant \dfrac{1}{2}\sum\limits_{n=1}^{2}E_{f}(\Pi_{n}|\psi\rangle)=E_{f}(|\psi\rangle).
\end{equation}
It follows that for any pure state $\rho=|\psi\rangle\langle\psi|$,
\begin{equation}
E_{f}(\rho^{\text{PI}})\leqslant E_{f}(\rho).
\end{equation}
For any bipartite mixed quantum state $\rho=\sum\limits_{i}p_{i}|\psi_{i}\rangle\langle\psi_{i}|$, there is
\begin{equation}
\rho^{\text{PI}}=\dfrac{1}{2}\sum\limits_{n=1}^{2}\Pi_{n}\rho\Pi_{n}^{\dag}=\sum\limits_{i}p_{i}\left(\dfrac{1}{2}\sum\limits_{n=1}^{2}\Pi_{n}|\psi_{i}\rangle\langle\psi_{i}|\Pi_{n}^{\dag}\right)
=\sum\limits_{i}p_{i}(|\psi_{i}\rangle\langle\psi_{i}|)^{\text{PI}}.
\end{equation}
From the convexity of entanglement measure constructed by the convex roof method, we obtain
\begin{equation}
E_{f}(\rho^{\text{PI}})\leqslant \sum\limits_{i}p_{i}E_{f}\big((|\psi_{i}\rangle\langle\psi_{i}|)^{\text{PI}}\big)\leqslant \sum\limits_{i}p_{i}E_{f}(|\psi_{i}\rangle).
\end{equation}
This implies that for any mixed state $\rho$
\begin{equation}
E_{f}(\rho^{\text{PI}})\leqslant E_{f}(\rho).
\end{equation}
Since the above proof is valid for any choice of (local) bases, hence, we derive the inequality (\ref{observation}).

Next, we discuss the case of the entanglement measures which are not convex-roof extended measures by taking several entanglement measures as examples.

The \textit{relative entropy of entanglement} is a fundamental entanglement measure, as the relative entropy is an important function in quantum information theory \cite{0004045,RMP74.197}. It is defined as
\begin{equation}
E_{R}(\rho):=\min\limits_{\sigma\in \mathcal{M}_{S}}S(\rho||\sigma),
\end{equation}
where $\mathcal{M}_{S}$ is the set of separable states and $S(\rho||\sigma):=\text{tr}[\rho\log_{2}\rho]-\text{tr}[\rho\log_{2}\sigma]$ is the quantum relative entropy.

Let $\sigma'$ is the closest separable state to $\rho$, that is, $E_{R}(\rho)=S(\rho||\sigma')\leqslant S(\rho||\sigma)$ for any $\sigma\in \mathcal{M}_{S}$. By using of unitary invariance of quantum relative entropy, we have $S(\rho||\sigma')=S(\Pi_{2}\rho\Pi_{2}^{\dag}||\Pi_{2}\sigma'\Pi_{2}^{\dag})\leqslant S(\Pi_{2}\rho\Pi_{2}^{\dag}||\Pi_{2}\sigma\Pi_{2}^{\dag})=S(\rho||\sigma)$. Denote $\Pi_{2}\mathcal{M}_{S}\Pi_{2}^{\dag}=\{\Pi_{2}\sigma\Pi_{2}^{\dag}|\sigma\in \mathcal{M}_{S}\}$, it is easy to see that $\Pi_{2}\mathcal{M}_{S}\Pi_{2}^{\dag}=\mathcal{M}_{S}$, and $E_{R}(\Pi_{2}\rho\Pi_{2}^{\dag})=S(\Pi_{2}\rho\Pi_{2}^{\dag}||\Pi_{2}\sigma'\Pi_{2}^{\dag})$. Then,

\begin{equation}
\begin{aligned}
E_{R}(\rho^{\text{PI}})&\leqslant S\left(\rho^{\text{PI}}\Big|\Big|\frac{\sigma'+\Pi_{2}\sigma'\Pi_{2}^{\dag}}{2}\right)\\
&\leqslant \frac{1}{2}S(\rho||\sigma')+\frac{1}{2}S(\Pi_{2}\rho\Pi_{2}^{\dag}||\Pi_{2}\sigma'\Pi_{2}^{\dag})\\
&=S(\rho||\sigma'),
\end{aligned}
\end{equation}
where joint convexity of quantum relative entropy \cite{PRL30.434,JMP1973}, i.e., for any $p\in [0,1]$ and any four states $\rho_{1}, \rho_{2}, \sigma_{1}, \sigma_{2}$, $S\big(p\rho_{1}+(1-p)\rho_{2}||p\sigma_{1}+(1-p)\sigma_{2}\big)\leqslant pS(\rho_{1}||\sigma_{1})+(1-p)S(\rho_{2}||\sigma_{2})$, has been used in the second inequality.

The \textit{negativity} is defined both for pure and mixed states as \cite{PRA58.883,PRA65.032314}
\begin{equation}
\mathcal{N}(\rho):=\frac{\|\rho^{T_{A}}\|_{1}-1}{2}\qquad \left(=\frac{\| \rho^{T_{B}}\|_{1}-1}{2}\right),
\end{equation}
where $\|X\|_{1}:=\text{tr}\sqrt{X^{\dag}X}$ denotes the trace norm (i.e., the sum of all singular values), and $\rho^{T_{A}}$ ($\rho^{T_{B}}$) is the partial transposition of $\rho$ with respect to the subsystem $A$ ($B$).

Note that $\|(\Pi_{2}\rho\Pi_{2}^{\dag})^{T_{A}}\|_{1}=\|\rho^{T_{B}}\|_{1}=\|\rho^{T_{A}}\|_{1}$, then by the triangle inequality of the norm we have
\begin{equation}
\begin{aligned}
\|(\rho^{\text{PI}})^{T_{A}}\|_{1}&=\left\|\left(\frac{1}{2}\rho+\frac{1}{2}\Pi_{2}\rho\Pi_{2}^{\dag}\right)^{T_{A}}\right\|_{1}\\
&\leqslant \frac{1}{2}\|\rho^{T_{A}}\|_{1}+\frac{1}{2}\left\|\left(\Pi_{2}\rho\Pi_{2}^{\dag}\right)^{T_{A}}\right\|_{1}\\
&=\|\rho^{T_{A}}\|_{1}.
\end{aligned}\label{negativity}
\end{equation}
It follows that $\mathcal{N}(\rho^{\text{PI}})\leqslant \mathcal{N}(\rho)$, which implies the desired result.

The \textit{logarithmic negativity} is defined as \cite{PRL90.027901}
\begin{equation}
\mathcal{LN}(\rho):=\log_{2}\|\rho^{T_{A}}\|_{1}.
\end{equation}

Due to the monotonicity of the function $f(x)=\log_{2}x$ and inequality (\ref{negativity}), one can derive the desired result naturally.

Recently, a new entanglement measure, \textit{logarithmic convex-roof extended negativity}, is presented \cite{2007.09573}. It is defined as follows
\begin{equation}
E_{\mathcal{N}}(\rho):=\log_{2}[\mathcal{N}_{f}(\rho)+1],
\end{equation}
where $\mathcal{N}_{f}$ is the convex-roof extended negativity \cite{PRA68.062304}.

Combining the monotonicity of the function $f(x)=\log_{2}x$ and Observation, it is straightforward to give result.
\qed
\section{Conclusion}
In this paper, we have investigated the refinement of axiomatic approach of quantifying entanglement. A reasonable requirement for valid entanglement measures was studied in detail for bipartite quantum systems. We verify that  most known entanglement measures of bipartite quantum systems satisfy the new  requirement for (multipartite) entanglement measures \cite{PRL112.180501}, include entanglement
of formation, geometric measure of entanglement,  generalized entanglement concurrence, the relative entropy of entanglement, the negativity, convex-roof extended negativity, the logarithmic negativity and the logarithmic convex-roof extended negativity, etc.  
 It would be interesting to broaden our approach to the general multipartite entanglement measures, since this may contribute to a further understanding of entanglement properties of multipartite quantum systems.

\begin{acknowledgments}
This work was supported by the National Natural Science Foundation of China under Grant Nos: 12071110 and 11947073; the Hebei Natural Science Foundation of China under Grant Nos: A2020205014 and A2018205125, and the Education Department of Hebei Province Natural Science Foundation under Grant No: ZD2020167.
\end{acknowledgments}

\end{document}